\documentclass[
    ,final            % use final for the camera ready runs
  ]
  {aipproc}

\layoutstyle{8x11double}

\begin{document}

\title{High-pressure Debye-Waller and Gr\"uneisen parameters of Au and Cu}

\author{Matthias J. Graf}{
  address={Los Alamos National Laboratory, Los Alamos, New Mexico 87545, USA}
}

\author{Carl W. Greeff}{
  address={Los Alamos National Laboratory, Los Alamos, New Mexico 87545, USA}
}

\author{Jonathan C. Boettger}{
  address={Los Alamos National Laboratory, Los Alamos, New Mexico 87545, USA}
}

\begin{abstract}
The lattice vibrations are determined in the quasi-harmonic approximation for
elemental Au and Cu 
to twice their normal density by first-principles electronic 
band-structure calculations. It is found for these materials that the important
moments of the phonon density of states can be obtained to high accuracy from 
short-ranged force constant models. We discuss the implications for the 
Gr\"uneisen parameters on the basis of calculated phonon moments and their
approximations by using bulk moduli and Debye-Waller factors.

\bigskip
\noindent
{\small PACS: 63.30.+t, 91.60.Fe, 62.50.+p \hfill LA-UR:03-5245}
\end{abstract}

\maketitle

%%%%%%%%%%%%%%%%%%%%%%%%%%%%%%%%%%%%%%%%%%%%
%% MAINMATTER
%%%%%%%%%%%%%%%%%%%%%%%%%%%%%%%%%%%%%%%%%%%%

%\section{Introduction}

Accurate and reliable high-pressure standards are fundamental for the study of
matter under extreme conditions and in the earth sciences. With the advent 
of neutron and x-ray diffraction measurements up to several hundred GPa in 
diamond anvil cells it becomes important to be able to determine
precisely the pressure. 
At high pressures the equation of state (EOS) of elemental metals are 
commonly used as pressure scales. However, these scales are based on either 
extrapolation of low-pressure data or reduction from the Hugoniot to room 
temperature. 
The reduction of the Hugoniot curve onto $P-V$ isotherms requires model 
assumptions about the Gr\"uneisen parameter, which are experimentally
not well founded. 
Our ability to calculate accurate moments of the phonon distribution 
from first-principles electronic structures enables us to strongly constrain 
the Gr\"uneisen parameter and to establish an accurate EOS 
and high-pressure standard for elemental gold and copper.
This is an important step toward high-precision high-pressure experiments.

Since recent experiments raised concerns about the consistency of these 
standards \cite{Akahama2002}, we revisited the problem of high-pressure 
standards using a semi-empirical approach combined with first-principles
electronic structure calculations \cite{Greeff2003}. 
This method is different from the tight-binding approach that was 
earlier applied to solid copper \cite{Rudin2002}.
We determined the lattice vibrational contribution to the EOS
in the quasi-harmonic approximation of elemental {\it face centered cubic} 
Au and Cu by first-principles electronic structure calculations.
We found that the important moments of the phonon density of states can 
be obtained to high accuracy from short-ranged force constant models.

In this paper, we present the implications for the Gr\"uneisen parameter 
based on those calculations and their approximations by using
bulk moduli and Debye-Waller factors which, in principle, can be obtained
from ultrasound and neutron/x-ray diffraction measurements.
 The accuracy of our results is based on the accuracy of first-principles 
electronic structure calculations, which were performed in the 
local-density approximation (LDA) and generalized-gradient approximation (GGA)
of density functional theory.
  From these results elastic moduli and zone boundary phonon frequencies 
were obtained. Fitting simultaneously a short-ranged, second 
nearest-neighbor Born-von K\'arm\'an (BvK) force matrix model 
to the zone boundary frequencies and the elastic 
moduli enabled us to generate phonon dispersion curves in the entire 
Brillouin zone and to compute any phonon moment needed, which is typically 
accurate within a few percent.

\begin{table}
\begin{tabular}{llllllll}
\hline
  &
  & {$B$ \ [GPa]}
  & {$\omega_0/2\pi$ \ [THz]}
  & {$\langle u^2 \rangle$ \ [pm$^2$]}
  & {$\gamma$} 
  & {$\gamma_B$} 
  & {$\gamma_{DW}$} \\
\hline
Au & theo. & 167.7 & 3.49 & 83.2 & 2.99 & 2.76 & 3.03\\
   & expt. & 167 \ Ref. \cite{deLaunay} & 3.65 \ Ref. \cite{Dederichs} & 63 \ Ref. \cite{Killean} & 2.95 \ Ref. \cite{Wallace} &  & \\
Cu & theo. & 149.5 & 6.86 & 61.7 & 1.85 & 2.31 & 1.69\\
   & expt. & 137 \ Ref. \cite{deLaunay} & 6.43 \ Ref. \cite{Dederichs} & 76 \ Ref. \cite{IntTables} & 2.02 \ Ref. \cite{Wallace} &  & \\
\hline
\end{tabular}

\caption{Bulk modulus $B$, log-moment $\omega_0$, thermal mean-square 
 displacement $\langle u^2 \rangle$, and Gr\"uneisen parameter $\gamma$
 for Au and Cu at ambient conditions were obtained from the theoretical data in 
 Figs.  1(a) and 2(a) using Eq.~(\ref{gamma_V}). Experimental results are 
 listed where available.
}
\label{table1}
\end{table}

%\section{Gr\"uneisen parameter}
\smallskip

At high temperatures, i.e., in the classical limit, 
the temperature dependence of the pressure of a solid is dominated by the 
contribution from lattice vibrations, $P^{\rm vib}$.
Its temperature derivative at constant volume,
$(\partial P^{\rm vib}/\partial T)_V$, 
is proportional to the Gr\"uneisen parameter $\gamma$, 
which is defined by
\begin{equation}
\gamma = - \sum_{\bf k} \frac{d \ln \, \omega_{\bf k}}{d \ln \, V}
= - \frac{d \ln \, \omega_0}{d \ln \, V} 
\ ,
\label{gamma}
\end{equation}
where the phonon  frequencies $\omega_{\bf k}$ are functions of volume
only and the summation is over all eigenmodes.
Eq.~(\ref{gamma}) defines the logarithmic phonon moment $\omega_0$.

For practical reasons one often uses an interpolating Debye 
phonon model for calculating the EOS or for analyzing diffraction data,
instead of the more elaborate lattice dynamical models.
Since in a Debye phonon model the Debye frequency $\omega_D$ is identical
to all other phonon moments, the approximation
$\omega_D \approx \omega_n$, with $n \ge -3$,
is widely used for computing lattice vibrational properties
and Gr\"uneisen parameters. 
The moments $\omega_{-2}$ and $\omega_{-3}$ are related to the Debye-Waller
factor at high temperatures and the sound velocity, respectively
\cite{Wallace}.
Using the theoretical relationships for the thermal mean-square
displacement of an atom, 
$\langle u^2 \rangle \propto T/\omega_{-2}^2$, and the sound speed, 
$c \propto \omega_{-3} \propto \sqrt{B V^{1/3}}$,
we find the following
approximations to the Mie-Gr\"uneisen theory for the high-temperature 
Gr\"uneisen parameter,
%$\gamma \approx \gamma_D = - \frac{d \ln \, \omega_D}{d \ln \, V}$,

\begin{eqnarray}
\gamma &\approx&  \gamma_B = -\frac{1}{6} - \frac{1}{2} 
	\frac{d \ln \,B}{d \ln \,V} 
\ ,
\label{gamma_B}
\\
\gamma &\approx&  \gamma_{DW} = \frac{1}{2}
	\frac{d \ln \,\langle u^2 \rangle}{d \ln \,V} 
\ .
\label{gamma_DW}
\end{eqnarray}

The derivation of Eq.~({\ref{gamma_B})  requires a constant Poisson ratio
(Slater approximation). 
Moruzzi et al.  \cite{Moruzzi} studied extensively its application
to the 4d transition elements and found good agreement with experiment 
at ambient conditions.
The advantage of using Eq. (\ref{gamma_B}) for estimating the Gr\"uneisen
parameter is that it is readily accessible from 
ultrasound and diffraction measurements.
On the other hand, the expression for the thermal parameter 
in Eq.~(\ref{gamma_DW}) has not seen much application, 
because of the extreme difficulties of measuring
accurate thermal mean-square displacements in high-pressure diffraction
experiments \cite{Zhao}.
Since these different approximations of the Gr\"uneisen parameter
emphasize different phonon frequencies in the Brillouin zone compared
to the log-moment $\omega_0$, we do expect to find deviations from $\gamma$
by at least several percent, when using these approximate formulas,
reflecting the differences between different phonon moments.

\begin{table}
\begin{tabular}{lcccccccccc}
\hline
  & {$\gamma_{\rm ref}$}
  & {$\gamma(V_0)$}
  & {$\gamma(V_0)_{B}$}
  & {$\gamma(V_0)_{DW}$}
  & {$q$}
  & {$q_{B}$}
  & {$q_{DW}$}
  & {$\omega_0/2\pi$ \ [THz]}
  & {$B$ \ [GPa]}
  & {$\langle u^2 \rangle$ \ [pm$^2$]}
\\
\hline
Au & 2.95 & 2.95 & 2.72 & 3.00 & 1.229 & 1.064 & 1.481 & 3.48 & 166.9 & 83.8\\
Cu & 2.02 & 1.85 & 2.29 & 1.68 & 0.445 & 0.774 & 0.623 & 6.86 & 149.1 & 61.3\\
\hline
\end{tabular}

\caption{Fit parameters and $\omega_0$, $B$, and $\langle u^2 \rangle$
  for Au and Cu at ambient conditions were obtained from fitting the theoretical
  data in Figs. 1(a) and 2(a) using Eq.~(\ref{gamma_q}).
  The reference values $\gamma_{\rm ref}$ are from \cite{Wallace}
  and were derived using the thermodynamic definition of 
  $\gamma = (V/C_V) ({\partial P^{\rm vib}}/{\partial T})_V$, with the 
  specific heat $C_V$.
}
\label{table2}
\end{table}

%\section{Results and Discussion}
\smallskip

The calculation of the phonon moments $\omega_n$ requires knowledge of the
phonon frequencies for all {\bf k} points in the Brillouin zone 
\cite{Wallace}. The direct first-principles calculation of
frequencies on a dense {\bf k} mesh is computationally intensive, 
while the phonon dispersions of most elements can be easily
parameterized using lattice dynamical models like a generalized 
BvK force matrix model. Often it suffices to use a short-ranged
force model to compute the low order moments accurately within a few percent
\cite{Graf2003}.  In particular, for elemental
Au and Cu the log-moment $\omega_0$ is
converged to less than 1\% with a 2nd nearest-neighbor interatomic shell
model at ambient conditions.
Thus, we chose to calculate four zone boundary phonon frequencies corresponding
to the transverse and longitudinal eigenmodes at the $X$ and $L$ points of the
Brillouin zone. These are computed with standard frozen-phonon methods.
Additionally, three elastic moduli are computed using the method by
S\"oderlind et al. \cite{Soederlind}. We fitted these results to a 2nd 
nearest-neighbor BvK force model, which then
allows the evaluation of the frequencies $\omega_{\bf k}$ in the entire
Brillouin zone. Details of this calculation will be published elsewhere
\cite{Greeff2003}.

We fitted the theoretical $\omega_0$, $B$, and $\langle u^2 \rangle$ 
results to a functional form that gives a realistic $V$-dependence of the 
Gr\"uneisen parameter, which has been used in many EOS
calculations \cite{Abdallah},
\begin{equation}
\gamma(V) = \gamma^\infty + A_1 \left( {V}/{V_0} \right)
	+ A_2 \left( {V}/{V_0} \right)^2
\ ,
\label{gamma_V}
\end{equation}
where $V_0$ is the volume at ambient pressure, $\gamma^\infty$ is the 
infinite density limit of $\gamma$, and $A_1$ and $A_2$ are fit parameters. 
Recently, it has been argued for the value
$\gamma^\infty = 1/2$, instead of the commonly used 
$\gamma^\infty = 2/3$ \cite{Burakovsky}. 
However, our results are insensitive to this
difference and hence we chose $\gamma^\infty = 2/3$.
The corresponding fitted values are listed in Table \ref{table1}.

In order to test the robustness of the calculated $\gamma$ values using
Eq.~(\ref{gamma_V}), we also fitted $\omega_0$, $B$, and
$\langle u^2 \rangle$ to the widely used expression
\begin{equation}
\gamma(V) = \gamma(V_0) \, \left( {V}/{V_0} \right)^q
\ ,
\label{gamma_q}
\end{equation}
where the fit parameters $\gamma(V_0)$ and $q$ are listed in 
Table \ref{table2}; see also the inserts of Figs. 1(b) and 2(b). 

In Figs.~(1) and (2) we show the normalized log-moments, bulk moduli,
and thermal mean-square displacement parameters 
of elemental Au and Cu based on our first-principles calculations.
To emphasize the very similar scaling behavior of these lattice dynamical
properties, we normalized their values by their corresponding ambient
condition values.

We used the functional forms for the Gr\"uneisen parameter given in 
Eqs.~(\ref{gamma_V}) and (\ref{gamma_q}) to integrate Eqs.
(\ref{gamma}) through (\ref{gamma_DW}). This allowed us to fit the
theoretical log-moments, bulk moduli, and thermal parameters and to
extract the fitting parameters necessary for calculating the
corresponding Gr\"uneisen parameters shown in Figs.~(1b) and (2b).

\begin{figure}
 \includegraphics[angle=0, width=.43\textwidth]{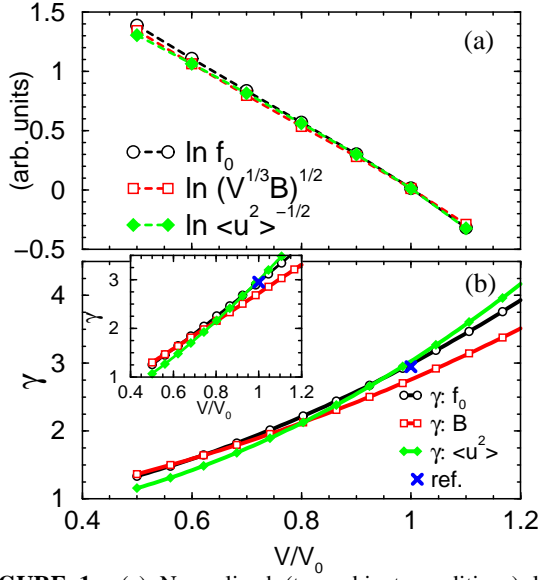}
 \caption{(a) Normalized (to ambient conditions) 
 log-moment, bulk modulus, and thermal mean-square displacement
 of Au at $T = 296 \ {\rm K}$ from electronic structure calculations at
 $V/V_0 = 0.5, 0.6, 0.7, 0.8, 0.9, 1.0, 1.1$.
 (b) Corresponding Gr\"uneisen parameters using Eq.~(\ref{gamma_V}).
 $\gamma(V_0) = 2.95$ \cite{Wallace} (cross) is shown for reference.
 Insert: Corresponding Gr\"uneisen parameters using Eq.~(\ref{gamma_q}). 
 }
\end{figure}

\begin{figure}
 \includegraphics[angle=0, width=.43\textwidth]{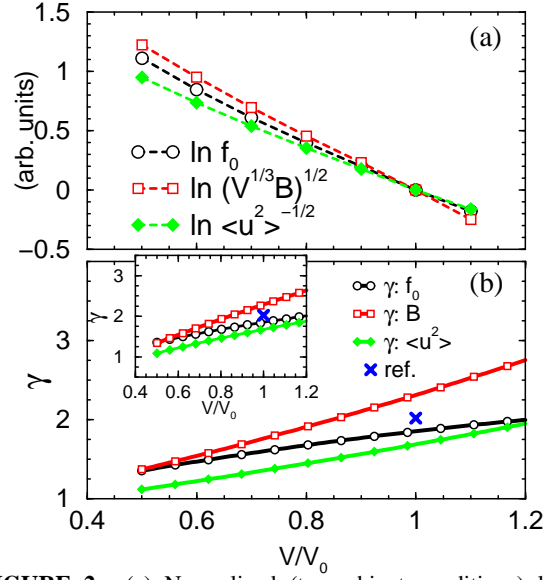}
 \caption{(a) Normalized (to ambient conditions)
 log-moment, bulk modulus, and thermal mean-square displacement 
 of Cu at $T = 296 \ {\rm K}$ from electronic structure calculations at 
 $V/V_0 = 0.5, 0.6, 0.7, 0.8, 0.9, 1.0, 1.1$.
 (b) Corresponding Gr\"uneisen parameters using Eq.~(\ref{gamma_V}).
 $\gamma(V_0) = 2.02$ \cite{Wallace} (cross) is shown for reference.
 Insert: Corresponding Gr\"uneisen parameters using Eq.~(\ref{gamma_q}).
 }
\end{figure}

We found for the volume compressions considered here that $\gamma$ of Au 
and Cu are quite well described by either expression (\ref{gamma_V}) or 
(\ref{gamma_q}). The differences in the calculated $\gamma$ values are 
only a few percent. A detailed discussion of the accuracy of the different
functional forms for $\gamma(V)$ and its consequences for the equation of
state and the Hugoniot is given in Ref. \cite{Greeff2003}.
In the case of Au, the approximate expressions for the Gr\"uneisen
parameter in Eqs. (\ref{gamma_B}) and (\ref{gamma_DW}) give good agreement with
the correct high-temperature $\gamma$ obtained from Eq. (\ref{gamma}). 
The deviations of $\gamma_{B}$
and $\gamma_{DW}$ from $\gamma$ are mostly less than 8\%.
In the case of Cu, the deviations of $\gamma_{B}$ and $\gamma_{DW}$ 
from $\gamma$ are generally bigger than for Au, but never more than
15\% for compressions in the range $0.5 < V/V_0 < 1.0$. 
A possible explanation of the larger deviations for Cu may be that
the Debye temperature and equivalently the log-moment of Cu are almost
twice as large as for Au. Since the typical temperature of lattice vibrations
of Cu, $\hbar\omega_0 / k_B \approx 330\ {\rm K}$,
is slightly above room temperature ($300\ {\rm K}$), 
one may have to go beyond the classical approximation for accurately
calculating the Gr\"uneisen parameter at ambient conditions.

Comparing $\gamma_{DW}$ and $\gamma_B$ with $\gamma$, we find in the 
range of compressions $0.8 < V/V_0 < 1.0$ that the Debye-Waller approximation
$\gamma_{DW}$ results generally in better agreement with $\gamma$ than 
the approximation using the bulk modulus $\gamma_B$. 
This situation is reversed below $V/V_0 \sim 0.8$. 
At such high compression $\gamma_B$ is in very good agreement with $\gamma$
of Au and in good agreement for Cu,
while $\gamma_{DW}$ deviates the most.
A simple explanation of this very different behavior of $\gamma_B$ and
$\gamma_{DW}$ at high compression, i.e., below $V/V_0 \sim 0.8$, is
due to the drastic stiffening of the phonon frequencies and the simultaneous
increase of the bulk modulus with decreasing $V$. At such high pressures
and room temperature the excited phonons probe mostly the linear part of
the phonon dispersion. 
The slope of the dispersion near the zone center is crudely proportional 
to the bulk modulus, while the Debye-Waller factor averages
all frequencies weighted by the temperature dependent occupation factor
of each mode. Therefore, we expect Eq. (\ref{gamma_DW}) to fail when the
temperature becomes comparable to the high-temperature Debye-Waller 
phonon moment. Roughly speaking, for $T$ of the order of the log-moment,
$k_B T \sim \hbar\omega_0$.

%\section{Conclusions}
\smallskip

In summary, we have successfully computed with high accuracy
phonon moments of elemental gold and copper from first-principles
electronic structure calculations combined with a Born-von K\'arm\'an
force model.  From the logarithmic phonon moments we calculated the 
volume dependence of the Gr\"uneisen parameter up to twice the normal 
density of Au and Cu at ambient conditions.
Comparing the Gr\"uneisen parameter with approximations based on
the bulk modulus and the Debye-Waller factor, we found that for low
compression, $0.8 < V/V_0 < 1.0$, the approximation using
the thermal mean-square displacement is more accurate than the one using 
the bulk modulus, while at higher compression the bulk modulus 
gives generally better agreement with $\gamma$.
Therefore, a combination of $\gamma_B$ and $\gamma_{DW}$, which can
be obtained from ultrasound and diffraction measurements, 
can give an estimate of the volume dependence of the Gr\"uneisen
parameter $\gamma$ within approximately 10\%.
This provides a useful alternative for determining 
the Gr\"uneisen parameter besides using the thermodynamic relation, 
which depends on the knowledge of the bulk modulus, thermal expansion, 
and specific heat.

%%%%%%%%%%%%%%%%%%%%%%%%%%%%%%%%%%%%%%%%%%%%%%%%
%% BACKMATTER
%%%%%%%%%%%%%%%%%%%%%%%%%%%%%%%%%%%%%%%%%%%%%%%%

\bigskip

%\begin{theacknowledgments}
This work was supported by the Los Alamos National Laboratory
under the auspices of the U.S. Department of Energy.
%\end{theacknowledgments}

\bibliographystyle{aipproc}   % if natbib is available

\end{document}